\documentclass[letterpaper,oneside,openbib]{SprMixLaTeX}
\usepackage[sectionbib,longnamesfirst]{natbib}
\bibpunct{(}{)}{;}{a}{}{,}
\usepackage{graphicx,marvosym}    
\usepackage[lined,boxed,ruled]{algorithm2e}

\def\__{\char"5F} 

\begin{document}

\title*{Self-organizing Traffic Lights: A Realistic Simulation}

\author{Seung-Bae Cools\and Carlos Gershenson\and Bart D'Hooghe\Cross}

\institute{}


\maketitle

\vspace{0.7 cm}

\section{Introduction: Catch the Green Wave? Better Make Your Own!}

Everybody in populated areas suffers from traffic congestion
problems. To deal with them, different methods have been developed
to mediate between road users as best as possible. Traffic lights
are not the only pieces in this puzzle, but they are an important
one. As such, different approaches have been used trying to reduce
waiting times of users and to prevent traffic jams. The most
common consists of finding the appropriate phases and periods of
traffic lights to quantitatively optimize traffic flow. This
results in ``green waves" that flow through the main avenues of a
city, ideally enabling cars to drive through them without facing a
red light, as the speed of the green wave matches the desired
cruise speed for the avenue. However, this approach does not
consider the current state of the traffic. If there is a high
traffic density, cars entering a green wave will be stopped by
cars ahead of them or cars that turned into the avenue, and once a
car misses the green wave, it will have to wait the whole duration
of the red light to enter the next green wave. On the other hand,
for very low densities, cars might arrive too quickly at the next
intersection, having to stop at each crossing. This method is
certainly better than having no synchronization at all, however,
it can be greatly improved.

Traffic modelling has enhanced greatly our understanding of this
complex phenomenon, especially during the last
decade \citep%
{PrigogineHerman1971,Traffic95,Traffic97,Traffic99,Helbing1997,HelbingHuberman1998},
suggesting different improvements to the traffic infrastructure.
One of these consists of adapting the traffic lights to the
current traffic conditions. Indeed, modern ``intelligent" advanced
traffic management systems (ATMS) use learning methods to adapt
phases of traffic lights, normally using a central computer
\citep{FHA1998,SCOOT1981}. The self-organizing approach we present
here does not need a central computer, as the global
synchronization is adaptively achieved by local interactions
between cars and traffic lights, generating flexible green waves
on demand.

We have previously shown in an abstract simulation
\citep{Gershenson2005} that self-organizing traffic lights can
greatly improve traffic flow for any density. In this paper, we
extend these results to a realistic setting, implementing
self-organizing traffic lights in an advanced traffic simulator
using real data from a Brussels avenue. In the next section, a
brief introduction to the concept of  self-organization is given.
The SOTL control method is then presented, followed by the moreVTS
simulator. In Section \ref{sec:results}, results from our
simulations are shown, followed by Discussion, Future Work, and
Conclusions.

\section{Self-organization}

The term \emph{self-organization} has been used in different areas with
different meanings, as is cybernetics \citep{vonFoerster1960,Ashby1962},
thermodynamics \citep{NicolisPrigogine1977}, biology \citep{CamazineEtAl2003},
mathematics \citep{Lendaris1964}, computing \citep{HeylighenGershenson2003},
information theory \citep{Shalizi2001}, synergetics \citep{Haken1981}, and
others \citep{SkarCoveney2003} (for a general overview, see \citep%
{Heylighen2003sos}). However, the use of the term is subtle, since any
dynamical system can be said to be self-organizing or not, depending partly
on the observer \citep{GershensonHeylighen2003a,Ashby1962}: If we decide to call a ``preferred" state or set of states (i.e. attractor) of a system ``organized", then the dynamics will lead to a self-organization of the system.

It is not necessary to enter into a philosophical debate on the theoretical
aspects of self-organization to work with it, so a practical notion will
suffice \citep{Gershenson2006}:

\begin{quotation}

A system \emph{described} as self-organizing is one in which
elements \emph{interact} in order to achieve \emph{dynamically} a global function or behavior.
\end{quotation}

This function or behavior is not imposed by one single or a few
elements, nor determined hierarchically. It is achieved
\emph{autonomously} as the elements interact with one another.
These interactions produce feedbacks that regulate the system. If
we want the system to solve a problem, it is useful to describe a
complex system as self-organizing when the ``solution" is not
known beforehand and/or is changing constantly. Then, the solution
is dynamically sought by the elements of the system. In this way,
systems can adapt quickly to unforeseen changes as elements
interact locally. In theory, a centralized approach could also
solve the problem, but in practice such an approach would require
too much time to compute the solution and would not be able to
keep the pace with the changes in the system and its environment.

In engineering, a self-organizing system would be one in which
elements are designed to \emph{dynamically}  and
\emph{autonomously}  solve a problem or perform a function at the
system level. Our traffic lights are self-organizing because each
one makes a decision based only on local information its own
state. Still, they manage to achieve robust and adaptive global
coordination.

\section{Self-organizing Traffic Lights: the Control Method}

In the SOTL method (originally named \textit{sotl-platoon} in
\cite {Gershenson2005}), each traffic light, i.e. intersection,
keeps a counter $\kappa _{i}$ which is set to zero when the light
turns red and then incremented at each timestep by the number of
cars approaching \emph{only} the red light (i.e. the next one a
car will reach) independently of the status or speed of the cars
(i.e. moving or stopped). When $\kappa _{i}$ (representing the
integral of cars over time) reaches a threshold $\theta $, the
green light at the same intersection turns yellow, and the
following time step it turns red with $\kappa _{i}=0$ , while the
red light which counted turns green. In this way, if there are
more cars approaching or waiting behind a red light, this will
turn into green faster than if there are only few cars. This
simple mechanism achieves self-organization in the following way:
if there are single or few cars, these will be stopped for more
time behind red lights. This gives time for other cars to join
them. As more cars join the group, cars will wait less time behind
red lights. With a sufficient number of cars, the red lights will
turn green even before they reach the intersection, generating
``green corridors". Having ``platoons" or ``convoys" of cars
moving together improves traffic flow, compared to a homogeneous
distribution of cars, since there are large empty areas between
platoons, which can be used by crossing platoons with few
interferences.

The following constraint is considered to prevent traffic lights from switching too fast when there are high densities: A traffic light will not be changed if the number of
time steps is less than a minimum phase, i.e. $\varphi _{i}<\varphi _{\min }$
($\varphi _{i}$ is the time since the light turned green).

Two further conditions are taken into account to regulate the size
of platoons. Before changing a red light to green, the controller
checks if a platoon is crossing through, in order not to break it.
More precisely, a red light is not changed to green if on the
crossing street there is at least one car approaching within  a
distance $\omega $ from the intersection. This keeps crossing
platoons together. For high densities, this condition alone would
cause havoc, since large platoons would block the traffic flow of
intersecting streets. To avoid this, we introduce a second
condition. Condition one is not taken into account if there are
more than $\mu $ cars approaching the green light. Like this, long
platoons can be broken, and the restriction only comes into place
if a platoon will soon be through an intersection.

The SOTL method is formally summarized in Algorithm \ref{alg:sotl}.

\IncMargin{0.5cm}
\RestyleAlgo{boxed}
\LinesNumbered
\begin{algorithm}[h!]
\begin{footnotesize}
\Indp

\ForEach{$(timestep)$}{
$\kappa_{i}$ += $cars_{approachingRed}$ in $\rho$\;
\If {($\varphi _{i}\geq \varphi _{\min }$)}{
 \If {\textbf{not} ($0 < cars_{approachingGreen}$ in $\omega < \mu $)}{
   \If {($\kappa_{i}$ $\geq $ $\theta $)}{
      $switchlight_{i}()$\;
      $\kappa_{i} = 0$
    }
  }
}
}

  \caption{Self-organizing traffic lights (SOTL) controller.}
\label{alg:sotl}
\end{footnotesize}
\end{algorithm}
\DecMargin{0.5cm}

This method has no phase or internal clock. If there are no cars
approaching a red light, the complementary one can stay green. We say that this method is self-organizing because the
global performance is given by the local rules followed by each traffic
light: they are ``unaware" of the state of other intersections and still manage
to achieve global coordination.

The method uses a similar idea to the one used by \citet{PorcheLafortune1998}, but with a much simpler
implementation. There is no costly prediction of arrivals at intersections,
and no need to establish communication between traffic lights to achieve
coordination and there are not fixed cycles.

\section{A Realistic Traffic Simulator: moreVTS}

Our simulator \citep{moreVTS} (A More Realistic Vehicle Traffic Simulator) is the third of a series of open source projects building on the previous one, developed in Java. Green Light District (GLD) was developed by the Intelligent Systems Group at the University of Utrecht \citep{gld-url,WieringEtAl2004}. Then, GLD was improved by students in Argentina within the iAtracos project, which we used as a starting point for our simulator, which introduces realistic physics into the simulation. Among other things, acceleration was introduced, and the scale was modified so that one pixel represents one meter and one cycle represents one second.

The simulator allows the modelling of complex traffic
configurations, enabling the user to create maps and then run
simulations varying the densities and types of road users.
Multiple-lane streets and intersections can be arranged, as well
as spawn and destination frequencies of cars. For implementation
details of moreVTS, the reader is referred to \citet{Cools2006}.

\begin{figure}[t]
\begin{center}
\includegraphics[width=12cm]{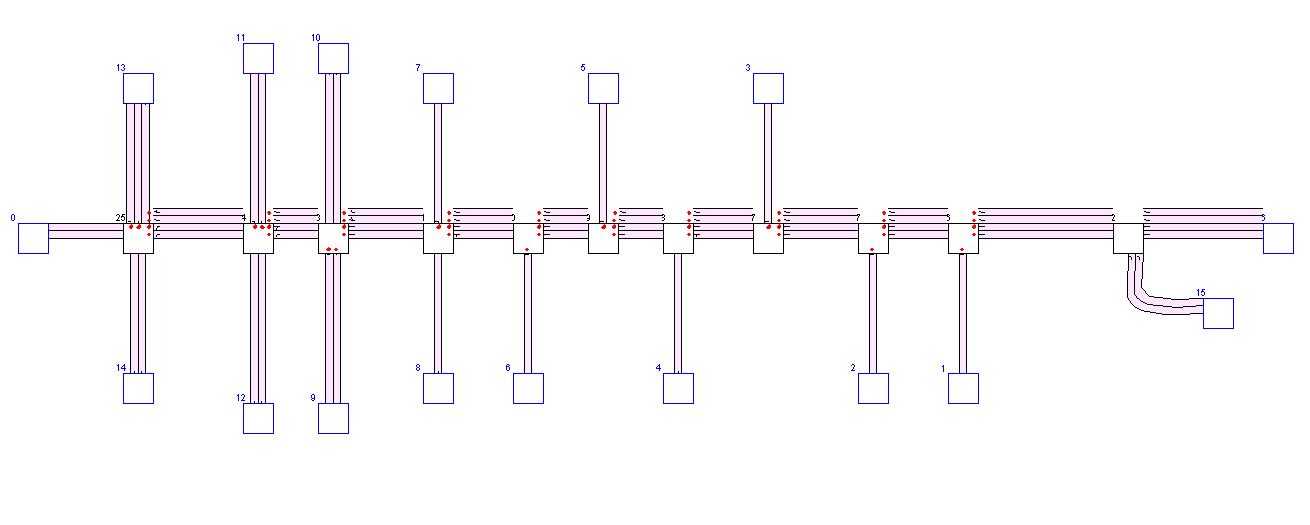}
\caption{Simulation of the Wetstraat and intersecting streets. Cars flow westward on the Wetstraat. Red dots represent traffic lights for each incoming lane at intersections.}
\label{cap:map3_wetstraat}
\end{center}
\end{figure}

The self-organizing traffic light controller described in the
previous section was implemented in moreVTS. Using data provided
by the Brussels Capital Region, we were able to build a detailed
simulation of the Rue de la Loi/Wetstraat, a four-lane westwards
one-way avenue in Brussels which gathers heavy traffic towards the
centre of the city. We used the measured average traffic densities
per hour on working days for 2004 (shown in Table
\ref{table:wetstraat_vehicle_count}) and the current ``green wave"
method, which has a period of 90 seconds, with 65 seconds for the
green phase on the Wetstraat, 19 for the green phase on side
streets, and 6 for transitions. This enabled us to compare our
self-organizing controller with a standard one in a realistic
setting. Figure \ref{cap:map3_wetstraat} shows the simulation view
of the Wetstraat and its surrounding streets.

\begin{table}[t]
\begin{center}
\begin{tabular}{|c|c|c|c|c|c|c|c|c|c|c|c|}
\hline
0 &1 & 2 & 3 & 4 & 5 & 6 & 7 & 8 & 9 & 10 & 11
\tabularnewline \hline
476 & 255 & 145 & 120 & 175 & 598 & 2933 & 5270 & 4141 & 4028 & 3543 & 3353
\tabularnewline \hline
\end{tabular}

\begin{tabular}{|c|c|c|c|c|c|c|c|c|c|c|c|}
\hline
12 & 13 & 14 & 15 & 16 & 17 & 18 & 19 & 20 & 21 & 22 & 23
\tabularnewline \hline
3118 & 3829 & 3828 & 3334 & 3318 & 3519 & 3581 & 3734 & 2387 & 1690 & 1419 & 1083
\tabularnewline \hline
\end{tabular}
\caption{Average vehicle count per hour at the beginning of the Wetstraat. Data kindly provided by the Brussels Capital Region}
\label{table:wetstraat_vehicle_count}
\end{center}
\end{table}

The data from Table \ref{table:wetstraat_vehicle_count} is for the
cars entering the Wetstraat on the East, so the spawn rates for
the two nodes in the simulation representing this were set
according to this data. For the other nodes, the spawn and
destination frequencies were set based on a field study we
performed in May 2006, comparing the percentage of cars that flow
through the Wetstraat and those that flow through side streets,
enter, or leave the Wetstraat. These percentages were kept
constant, so that when the density of cars entering the Wetstraat
changed, all the other spawn rates changed in the same proportion.
On average, for each five cars flowing through a side street, one
hundred flow through the Wetstraat. This is not the case of the
Kuststraat, a two way avenue at the West of the Wetstraat (second
and third crossing streets from left to right on Fig.
\ref{cap:map3_wetstraat}), where for 100 cars driving through the
Wetstraat, about 40 turn right, 40 turn left, and only 20 go
straight, while 20 more drive through the Kuststraat (about 10 in
each direction). The precise spawn rates and destination
frequencies are given in \citet[pp. 55--57]{Cools2006}.

\section{Results}
\label{sec:results}

To measure the performance of the current green wave method and
our self-organizing controller, we used the average trip waiting
times (ATWT). The trip waiting time for one car is the travel time
minus the minimum possible travel time (i.e. travel distance
divided by the maximum allowed speed, which for the Wetstraat simulation is about sixty seconds).

\begin{figure}[h]
\begin{center}
\includegraphics[angle=-90, width=1\textwidth]{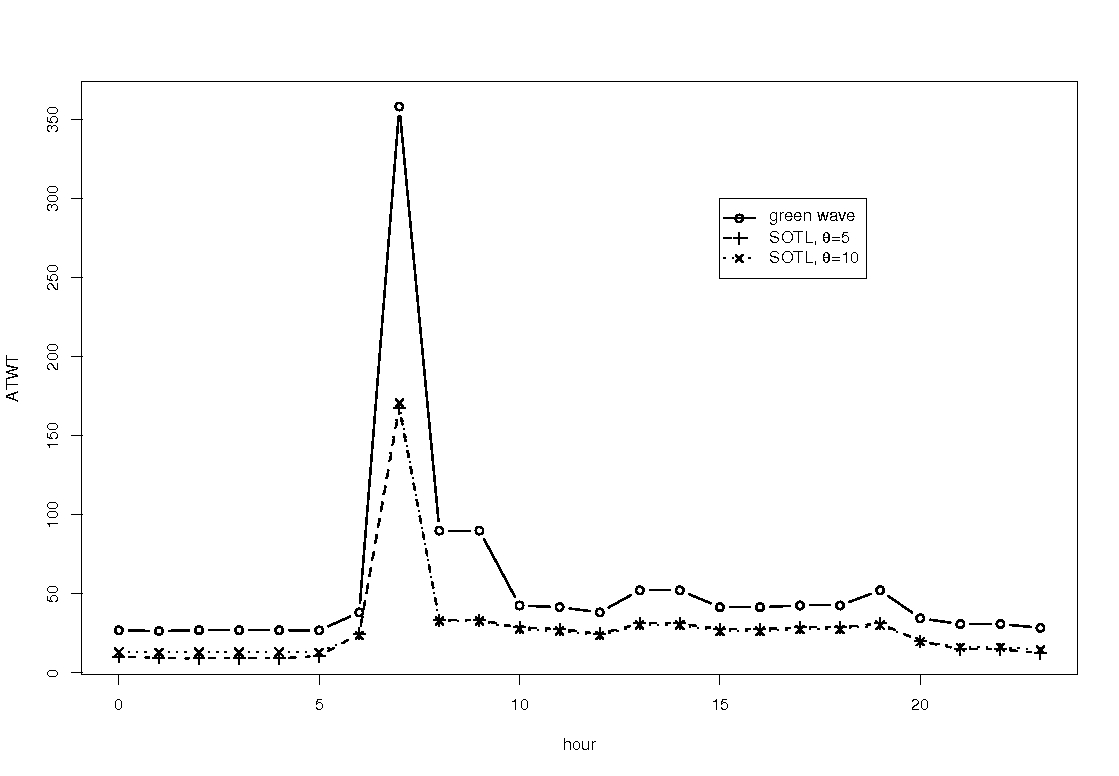}
\caption{Average trip waiting times (ATWT) at different hours of the day, green wave controller and SOTL controller
with $\protect\varphi _{min}=5$ and $\protect\theta =5;10$}
\label{fig:results_wetstraat}
\end{center}
\end{figure}

Several simulation runs were performed to find the best parameters for the SOTL method. For each parameter and traffic density, five simulation runs representing one hour, i.e. 3600 cycles, were averaged. The results were robust and consistent, with SOTL performing better than the green wave method for a wide range of parameters  $\theta $ and $ \varphi _{\min }$ \citep{Cools2006}. Only the best ones are shown in Fig. \ref{fig:results_wetstraat}, together with the results for the green wave method. The cruise speed used was 14 m/s, $\omega = 25$ and $ \mu = 3$. Since some densities from Table \ref{table:wetstraat_vehicle_count} are very similar, we averaged and considered the same densities for 2:00, 3:00 and 4:00;  8:00 and 9:00; 10:00, 17:00 and 18:00; 11:00, 15:00 and 16:00; 13:00, 14:00 and 19:00; and 21:00 and 22:00.

As Fig. \ref{fig:results_wetstraat} shows, there is considerable reduction in ATWT using SOTL instead of the current green wave method. The ATWT for the densities at different hours using SOTL were from 34\% to 64\% of the ATWT for the green wave method, and on average 50\%. Since the minimum travel time for the Wetstraat is about one minute, while the overall ATWT for the green wave method is also about one minute and for SOTL about half, the improvement in the average total travel times would be of about 25\%, i.e. cars under a green wave method would take 33\% more time to reach their destination than those under SOTL.
This shows with a realistic simulation that SOTL improves greatly traffic flow compared to the current green wave method.

\section{Discussion}

The green wave method works well for regulating traffic on the
Wetstraat, since most of the traffic flows through it. Still,
having no consideration about the actual state of the traffic has
several drawbacks. It can give a green light to a side street even
if there are no cars on it, or when a group of cars is about to
cross. Also, if the traffic density is high, the speed of the cars
will be slower than that of the green wave. And when a car misses
a green wave, it will have to wait a full cycle to enter the next
one.

 Having actual information about the traffic state enables SOTL to adapt to the current situation: it only gives green lights on demand, so time will not be wasted for streets without cars, and the street with more cars will have more demand, thus more green lights. Cars do have to wait behind red lights, but since while doing so they are demanding to cross, it is very unlikely that a car will have to wait more than $\phi_{min}$. Moreover, when a car is stopped, a platoon is likely to be formed, accelerating the switching of green lights.

Another advantage of platoons is that they reduce entropy in the
city, defined via the probability of finding a car in any part of
the city. If there is maximal entropy, there is the same
probability of finding a car anywhere on the city. This increases
the probability of interference, i.e. that two cars will meet at
an intersection, thus one needs to stop. The opposite extreme is
less desirable: if we have a certainty of the position of every
car, it is because they are stopped, i.e. in a traffic jam.
However, platoons offer a useful balance: there is a high
probability that a car will be close to another car, i.e. in a
group. Thus, there are many free spaces left between platoons,
that other platoons can exploit to cross without interferences.
There will be interferences, but these will be minimal.

\section{Future Work}

The following list summarizes the future work.

\begin{itemize}
\item A method similar to SOTL has been used successfully in the
United Kingdom for some time, but only for isolated intersections
\citep{VincentYoung1986}. Indeed, it is not obvious to expect that
traffic lights without direct communication would be able to
coordinate robustly. In any case, the technology to implement it
is already available, so a pilot study could be quickly deployed
in a real city. Since the traffic lights are adaptive, only a few
intersections could be changed, which would adapt to the control
method used in the rest of the city. This also would make it easy
to incrementally introduce them in large cities.

\item We have observed that there is a monotonic relation between
the best $\theta$ and the traffic density \citep{Cools2006}.
Exploring this relation better could allow us to set a variable
$\theta$ depending on the current traffic density measured by the
traffic lights. However, since SOTL performs very well for a broad
range of parameters, it does not require the calculation of
precise parameters. In other words, SOTL is not sensitive to small
changes in parameters, making it a robust method.

\item The SOTL method could also be used to give preference to
certain users, e.g. public transport or emergency vehicles.
Simply, a weight would be given to each vehicle in the count
$\kappa_i$, so that vehicles with preference would be able to
trigger green lights by themselves. They would be equivalent to a
platoon of cars, thus being seamlessly integrated into the system.
This might be a considerable improvement compared to current
methods where some vehicles (e.g. buses in London, trams in
Brussels) have preference and the rest of the users are neglected,
in some cases even when there are no preferred vehicles nearby.

\item The ``optimal" sizes of platoons, depending on different features of a city, is an interesting topic to research. The parameters of SOTL can be regulated to promote platoons of a certain size, so knowing which size should be aimed at would facilitate the parameter search.

\item It would be interesting to compare SOTL with the Dresden method \citep{HelbingEtAl2005,LaemmerEtAl2006}, which couples oscillators using self-organization, whereas SOTL has no internal phases nor clocks.

\end{itemize}

\section{Conclusions}

In this chapter we presented results showing that a
self-organizing traffic lights control method considerably
improves the traffic flow compared to the current green wave
method, namely reducing on average waiting times by half. These
results are encouraging enough to continue refining and exploring
similar traffic light controllers and to implement them in real
cities, starting with pilot studies. However, we would not like to
motivate further the use of cars with an efficient traffic
control, since this would increase traffic densities and pollution
even more. Any city aiming at improving its traffic flow should
promote in parallel alternative modes of transportation, such as
cycling, walking, car pooling, or using public transport.

\section{Epilogue}

After this chapter was initially published, the self-organizing method was improved with three more rules \citep{GershensonRosenblueth:2010}, now achieving close to optimal performance for all densities. This was evaluated with an elementary cellular automaton model of city traffic \citep{RosenbluethGershenson:2010}. This abstract model is useful for comparing methods against a theoretical optimum and for detecting phase transitions. Since the self-organizing method is close to the theoretical optimum, there is little which can be further improved in traffic light coordination. We have also generalized the algorithm to complex intersections \citep{GershensonRosenblueth:2011}, implementing the elementary cellular automaton model on an hexagonal grid. The green wave method performs even worse when more streets have to be coordinated, while the self-organizing method manages to scale to the increased complexity of the traffic light coordination. In one case, as vehicle density varied, six phase transitions were identified. In another case, ten phase transitions occurred. Currently we are planning a pilot study to test the effectiveness of the self-organizing method with real traffic.

Self-organization has also been applied to other aspects of urban mobility \citep{Gershenson:2011b}. One interesting example involves public transportation systems \citep{Gershenson:2011a}, where self-organization manages to achieve better than optimal performance. Self-organizing systems are promising not only for improving mobility, but for a broad variety of urban problems \citep{Gershenson:2013}. To contribute to this goal, a methodology for designing and controlling self-organizing systems has been developed \citep{GershensonDCSOS}.

\section{Acknowledgements}

We should like to thank the Ministerie van het Brussels Hoofdstedelijk Gewest for their support, providing the data for the Wetstraat.

\bibliographystyle{apalike}
\bibliography{carlos,sos,traffic}

\begin{thebibliography}{}

\bibitem[Ashby, 1962]{Ashby1962}
Ashby, W.~R. (1962).
\newblock Principles of the self-organizing system.
\newblock In Foerster, H.~V. and {Zopf, Jr.}, G.~W., editors, {\em Principles
  of Self-Organization}, pages 255--278, Oxford. Pergamon.

\bibitem[Camazine et~al., 2003]{CamazineEtAl2003}
Camazine, S., Deneubourg, J.-L., Franks, N.~R., Sneyd, J., Theraulaz, G., and
  Bonabeau, E. (2003).
\newblock {\em Self-Organization in Biological Systems}.
\newblock Princeton University Press, Princeton, NJ, USA.

\bibitem[Cools, 2006]{Cools2006}
Cools, S.~B. (2006).
\newblock A realistic simulation for self-organizing traffic lights.
\newblock Unpublished BSc Thesis, Vrije Universiteit Brussel.

\bibitem[{Federal Highway Administration}, 1998]{FHA1998}
{Federal Highway Administration} (1998).
\newblock {\em Traffic Control Systems Handbook}.
\newblock U.S. Department of Transportation.

\bibitem[Gershenson, 2005]{Gershenson2005}
Gershenson, C. (2005).
\newblock Self-organizing traffic lights.
\newblock {\em Complex Systems}, 16(1):29--53.

\bibitem[Gershenson, 2006]{Gershenson2006}
Gershenson, C. (2006).
\newblock A general methodology for designing self-organizing systems.
\newblock Technical Report 2005-05, ECCO.

\bibitem[Gershenson, 2007]{GershensonDCSOS}
Gershenson, C. (2007).
\newblock {\em Design and Control of Self-organizing Systems}.
\newblock CopIt Arxives, Mexico.
\newblock http://tinyurl.com/DCSOS2007.

\bibitem[Gershenson, 2011]{Gershenson:2011a}
Gershenson, C. (2011).
\newblock Self-organization leads to supraoptimal performance in public
  transportation systems.
\newblock {\em {PLoS ONE}}, 6(6):e21469.

\bibitem[Gershenson, 2012]{Gershenson:2011b}
Gershenson, C. (2012).
\newblock Self-organizing urban transportation systems.
\newblock In Portugali, J., Meyer, H., Stolk, E., and Tan, E., editors, {\em
  Complexity Theories of Cities Have Come of Age: An Overview with Implications
  to Urban Planning and Design}, pages 269--279. Springer, Berlin Heidelberg.

\bibitem[Gershenson, 2013]{Gershenson:2013}
Gershenson, C. (2013).
\newblock Living in living cities.
\newblock {\em Artificial Life}, In Press.

\bibitem[Gershenson and Heylighen, 2003]{GershensonHeylighen2003a}
Gershenson, C. and Heylighen, F. (2003).
\newblock When can we call a system self-organizing?
\newblock In Banzhaf, W., Christaller, T., Dittrich, P., Kim, J.~T., and
  Ziegler, J., editors, {\em Advances in Artificial Life, 7th European
  Conference, {ECAL} 2003 {LNAI} 2801}, pages 606--614, Berlin. Springer.

\bibitem[Gershenson and Rosenblueth, 2012a]{GershensonRosenblueth:2010}
Gershenson, C. and Rosenblueth, D.~A. (2012a).
\newblock Adaptive self-organization vs. static optimization: A qualitative
  comparison in traffic light coordination.
\newblock {\em Kybernetes}, 41(3):386--403.

\bibitem[Gershenson and Rosenblueth, 2012b]{GershensonRosenblueth:2011}
Gershenson, C. and Rosenblueth, D.~A. (2012b).
\newblock Self-organizing traffic lights at multiple-street intersections.
\newblock {\em Complexity}, 17(4):23--39.

\bibitem[GLD, 2001]{gld-url}
GLD (2001).
\newblock Green {L}ight {D}istrict.

\bibitem[Haken, 1981]{Haken1981}
Haken, H. (1981).
\newblock Synergetics and the problem of selforganization.
\newblock In Roth, G. and Schwegler, H., editors, {\em Self-Organizing Systems:
  An Interdisciplinary Approach}, pages 9--13, New York. Campus Verlag.

\bibitem[Helbing, 1997]{Helbing1997}
Helbing, D. (1997).
\newblock {\em Verkehrsdynamik}.
\newblock Springer, Berlin.

\bibitem[Helbing et~al., 2000]{Traffic99}
Helbing, D., Herrmann, H.~J., Schreckenberg, M., and Wolf, D.~E., editors
  (2000).
\newblock {\em Traffic and Granular Flow '99: Social, Traffic, and Granular
  Dynamics}, Berlin. Springer.

\bibitem[Helbing and Huberman, 1998]{HelbingHuberman1998}
Helbing, D. and Huberman, B.~A. (1998).
\newblock Coherent moving states in highway traffic.
\newblock {\em Nature}, 396:738--740.

\bibitem[Helbing et~al., 2005]{HelbingEtAl2005}
Helbing, D., L\"{a}mmer, S., and Lebacque, J.-P. (2005).
\newblock Self-organized control of irregular or perturbed network traffic.
\newblock In Deissenberg, C. and Hartl, R.~F., editors, {\em Optimal Control
  and Dynamic Games}, pages 239--274. Springer, Dordrecht.

\bibitem[Heylighen, 2003]{Heylighen2003sos}
Heylighen, F. (2003).
\newblock The science of self-organization and adaptivity.
\newblock In Kiel, L.~D., editor, {\em The Encyclopedia of Life Support
  Systems}. EOLSS Publishers, Oxford.

\bibitem[Heylighen and Gershenson, 2003]{HeylighenGershenson2003}
Heylighen, F. and Gershenson, C. (2003).
\newblock The meaning of self-organization in computing.
\newblock {\em IEEE Intelligent Systems}, pages 72--75.

\bibitem[Hunt et~al., 1981]{SCOOT1981}
Hunt, P.~B., Robertson, D.~I., Bretherton, R.~D., and Winton, R.~I. (1981).
\newblock {SCOOT}-a traffic responsive method of coordinating signals.
\newblock Technical report, TRRL.

\bibitem[L\"{a}mmer et~al., 2006]{LaemmerEtAl2006}
L\"{a}mmer, S., Kori, H., Peters, K., and Helbing, D. (2006).
\newblock Decentralised control of material or traffic flows in networks using
  phase-synchronisation.
\newblock {\em Physica A}, 363(1):39--47.

\bibitem[Lendaris, 1964]{Lendaris1964}
Lendaris, G.~G. (1964).
\newblock On the definition of self-organizing systems.
\newblock {\em Proceedings of the IEEE}, 52(3):324--325.

\bibitem[more{VTS}, 2006]{moreVTS}
more{VTS} (2006).
\newblock A more realistic vehicle traffic simulator.

\bibitem[Nicolis and Prigogine, 1977]{NicolisPrigogine1977}
Nicolis, G. and Prigogine, I. (1977).
\newblock {\em Self-Organization in Non-Equilibrium Systems: From Dissipative
  Structures to Order Through Fluctuations}.
\newblock Wiley, Chichester.

\bibitem[Porche and Lafortune, 1999]{PorcheLafortune1998}
Porche, I. and Lafortune, S. (1999).
\newblock Adaptive look-ahead optimization of traffic signals.
\newblock {\em Journal of Intelligent Transportation Systems}, 4(3):209--254.

\bibitem[Prigogine and Herman, 1971]{PrigogineHerman1971}
Prigogine, I. and Herman, R. (1971).
\newblock {\em Kinetic Theory of Vehicular Traffic}.
\newblock Elsevier, New York.

\bibitem[Rosenblueth and Gershenson, 2011]{RosenbluethGershenson:2010}
Rosenblueth, D.~A. and Gershenson, C. (2011).
\newblock A model of city traffic based on elementary cellular automata.
\newblock {\em Complex Systems}, 19(4):305--322.

\bibitem[Schreckenberg and Wolf, 1998]{Traffic97}
Schreckenberg, M. and Wolf, D.~E., editors (1998).
\newblock {\em Traffic and Granular Flow '97}, Singapore. Springer.

\bibitem[Shalizi, 2001]{Shalizi2001}
Shalizi, C.~R. (2001).
\newblock {\em Causal Architecture, Complexity and Self-Organization in Time
  Series and Cellular Automata}.
\newblock PhD thesis, University of Wisconsin at Madison.

\bibitem[Sk{\aa}r and Coveney, 2003]{SkarCoveney2003}
Sk{\aa}r, J. and Coveney, P.~V., editors (2003).
\newblock {\em Self-Organization: The Quest for the Origin and Evolution of
  Structure}. Phil. Trans. R. Soc. Lond. A 361(1807).
\newblock Proceedings of the 2002 {Nobel Symposium} on self-organization.

\bibitem[Vincent and Young, 1986]{VincentYoung1986}
Vincent, R.~A. and Young, C.~P. (1986).
\newblock Self optimising traffic signal control using microprocessors - the
  {TRRL} {MOVA} strategy for isolated intersections.
\newblock {\em Traffic Engineering and Control}, 27(7-8):385--387.

\bibitem[{von Foerster}, 1960]{vonFoerster1960}
{von Foerster}, H. (1960).
\newblock On self-organizing systems and their environments.
\newblock In Yovitts, M.~C. and Cameron, S., editors, {\em Self-Organizing
  Systems}, pages 31--50, New York. Pergamon.

\bibitem[Wiering et~al., 2004]{WieringEtAl2004}
Wiering, M., Vreeken, J., Veenen, J.~V., and Koopman, A. (2004).
\newblock Simulation and optimization of traffic in a city.
\newblock In {\em {IEEE} Intelligent Vehicles Symposium {(IV'04)}}, pages
  453--458. IEEE.

\bibitem[Wolf et~al., 1996]{Traffic95}
Wolf, D.~E., Schreckenberg, M., and Bachem, A., editors (1996).
\newblock {\em Traffic and Granular Flow '95}, Singapore. World Scientific.

\end{thebibliography}

\end{document}